\documentclass[10pt,final,journal,letterpaper,oneside,twocolumn]{IEEEtran}
\pdfoutput=1

\usepackage{cite}

\usepackage{graphicx}
\usepackage{epstopdf}
\graphicspath{{Figures/}{../Figures/}}

\usepackage[caption=false,font=footnotesize]{subfig}

\usepackage[cmex10]{amsmath}

\usepackage{booktabs}

\usepackage{url}

\begin{document}

\title{On the 5G Wireless Communications at the Low Terahertz Band}

\author{Turker~Yilmaz,~\IEEEmembership{Student~Member,~IEEE,}
				and~Ozgur~B.~Akan,~\IEEEmembership{Fellow,~IEEE}
\thanks{The authors are with the Next-Generation and Wireless Communications Laboratory (NWCL), Department of Electrical and Electronics Engineering, Koc University, Istanbul, Turkey (e-mail: {turkeryilmaz, akan}@ku.edu.tr).}}

\maketitle

\begin{abstract}
\boldmath
Initiation of fourth generation (4G) mobile telecommunication system rollouts fires the starting pistol for beyond 4G research activities. Whereas technologies enhancing spectral efficiency have traditionally been the solution to data rate and network capacity increase demands, due to the already advanced techniques, returns of the even more complicated algorithms hardly worth the complexity increase any longer. In addition, surging number of connected devices now enables operative use of short-range communication methods. Also considering the recently approved standards for the 60 gigahertz (GHz) industrial, scientific and medical radio band, in this paper the transmission windows around 300 GHz is proposed to be utilized for the fifth generation wireless communication systems. Motivations for the low end of the terahertz (THz) band are provided in accordance with market trends, and standardization activities in higher frequencies are listed. Unified mathematical expressions of the electromagnetic wave propagation mechanisms in the low-THz band are presented. THz band device technologies are outlined and a complete survey of the state-of-the-art low-THz band circuit blocks which are suitable for mass market production is given. Future research directions are specified before the conclusion of the paper.
\end{abstract}

\begin{IEEEkeywords}
5G mobile communication, submillimeter wave communication, submillimeter wave propagation, submillimeter wave technology, submillimeter wave circuits, communication standards.
\end{IEEEkeywords}

\section{Introduction}
\label{sec:intr}

The necessity for greater human mobility brings the need to obtain information on the move, and thus an increase in mobile data traffic. According to the most recent networking index by Cisco Systems, global mobile data traffic grew 74\% in 2015 and is expected to rise at a compound annual growth rate (CAGR) of 53\% from 2015 to 2020. Another important trend is the projected increase at the global mobile network connection speeds. Whereas that average downstream speed for smartphones grew nearly 26\% to 7.5 megabits per second (Mb/s) in 2015, the quantity is anticipated to reach 12.5 Mb/s by 2020, following a five year 11\% CAGR \cite{ciscoVNI-2015}.

Minimum technical performance requirements for the modern fourth generation (4G) mobile telecommunication systems were outlined in 2008 by the International Telecommunication Union (ITU) Radiocommunication Sector's (ITU-R) Report ITU-R M.2134. In the document, the minimum downlink (DL) peak spectral efficiency was defined as 15 b/s/Hz and operation in wider bandwidths up to 100 megahertz (MHz) was encouraged, thus setting the theoretical DL peak data rate at 1500 Mb/s \cite{ITU-2134}. However, spectral efficiency was defined assuming a 4$\times$4 multiple input multiple output (MIMO) antenna configuration, whereas current 4G products operate on 2-stream MIMO and 4-stream MIMO is envisioned only for the user equipment (UE) of the fifth generation (5G) wireless communication systems, for which non-coherent detection would also be preferred \cite{WXu-2014W, WSchober-2002W}.

As 4G systems, although currently represent just 14\% of global mobile connections, are already being laid out by mobile network operators, research on 5G is taking pace. Many traffic forecast and market reports are getting published to guide initial specification and standardization activities. Two key common conclusions of these studies are an expectation of thousandfold increase in overall wireless communication traffic volume within a decade, and the forthcoming machine-to-machine (M2M) communication boom. Tenfold of the traffic rise is attributed to the increase in M2M connections, which is estimated to grow from 0.6 to 3.2 billion by 2020 owing to a five year 38\% CAGR, and the rest caused by the rise in traffic per device \cite{ciscoVNI-2015, Nokia-2011}. Based on these expectancies, 5G systems' peak data rate is required to be at least on the order of 10 gigabits per second (Gb/s).

Accommodating the anticipated traffic growth requires total throughput rise together with the data rates. The principal methods to accomplish this are evident: Increasing the operation bandwidth or spectral efficiency, or reducing the signalling overhead. Because the subject is applicable to nearly all areas of wireless communications, spectral efficiency enhancement efforts have traditionally led the research. A direct consequence of these studies are the spectrally very efficient systems in operation presently, one of which being the Long Term Evolution-Advanced with coordinated multi-point \cite{WHuq-2012W}. Whereas orthogonal frequency division multiplexing (OFDM) is utilized by the majority of the broadband communication systems \cite{WJeremic-2011W}, filter bank multicarrier, despite its higher complexity, is investigated for future communication systems too, due to its better spectral efficiency \cite{Farhang-Boroujeny-2011}.

Device densification inherent to the continuously growing number of mobile-connected devices leads to shorter propagation paths, originating a favourable situation for the employment of wider bandwidths over higher carrier frequencies \cite{WBoronin-2014W}. Taking advantage of this, increasing the operation frequency to the low end of the terahertz (THz) band is proposed in this paper as a solution to the data rate and network capacity requirements of the future 5G wireless communication systems.

The remainder of this paper is organized as follows. Section \ref{sec:std} presents a brief overview of the standardization activities on low-THz band, namely the 300 gigahertz (GHz) spectrum, and Section \ref{sec:model} outlines its channel characteristics and available models. Section \ref{sec:device} provides state-of-the-art THz device technologies that operate in room temperature and hold promise for use in commercial 5G products. Key open research issues are identified in Section \ref{sec:iss} and the article concludes with recapitulating remarks.

\section{Standardization Activities}
\label{sec:std}

The need for new spectral resources has been addressed by choosing the 60 GHz as the new industrial, scientific and medical (ISM) radio band to be used for unlicensed wireless communications. In addition to the two wireless personal area network (WPAN) standards, ECMA-387\cite{ECMA387-2010} and IEEE 802.15.3c\cite{802.15.3c-2009}, which have been available since December 2008 and October 2009, respectively, the only wireless local area network (WLAN) standard for the 60 GHz, IEEE 802.11ad, was ratified in December 2012. To provide an initial sense of the potential gains obtainable by the increase in operation frequency, 802.11ad standard makes use of the 9 GHz of unlicensed bandwidth available in most parts of the world between 57 and 66 GHz, by defining channel bandwidths to be 2160 MHz. This ample channel lets single carrier waveforms to reach a maximum data rate of 4620 Mb/s with $\pi$/2 16-ary quadrature amplitude modulation (16-QAM), and OFDM waveforms to reach 6756.75 Mb/s using 64-QAM\cite{802.11ad-2012}. Detailed explanations of all three 60 GHz standards are available in the literature\cite{XYilmaz-2016X}.

According to ITU's latest Radio Regulations, Edition of 2015, frequencies up to 275 GHz are completely allocated to various services and stations, whereas frequency bands in the range of 275-1000 GHz are identified only for the passive service applications of radio astronomy, Earth exploration-satellite and space research. Therefore, the spectrum beyond 275 GHz is nearly uninhabited and at present obtainable by any valuable service, including wireless and mobile communications. In fact, relevant standardization activities began in 2008 with the formation of IEEE 802.15 WPAN Terahertz Interest Group (IG THz) whose focus was on the THz frequency bands between 275 and 3000 GHz. By July 2013, the efforts both within the group and industry reached to the adequate level to transform IG THz into a study group, SG 100G, with the aim of developing project authorization request (PAR) and five criteria documents addressing 40/100 Gb/s over beam switchable wireless point-to-point links. The PAR, which identifies data centers for the beam switched point-to-point applications and adds kiosk and fronthaul and backhaul intra-device communications \cite{WGuan-2016W} to the usage models, was approved in March 2014, resulting in the formation of the Task Group 3d (TG3d). IG THz is also retained for applications outside the scope of the TG3d.

\section{THz Band Channel Property and Models}
\label{sec:model}

The answer to electromagnetic (EM) wave propagation is available since 1865 through Maxwell's equations. However, by making use of the general properties of typical wireless communication scenarios, the complexity of solving four differential equations for a point in space and time can be greatly reduced via specific channel models which provide satisfactory approximations.

\begin{figure}[!t]
\centering
\includegraphics[width=3.5in]{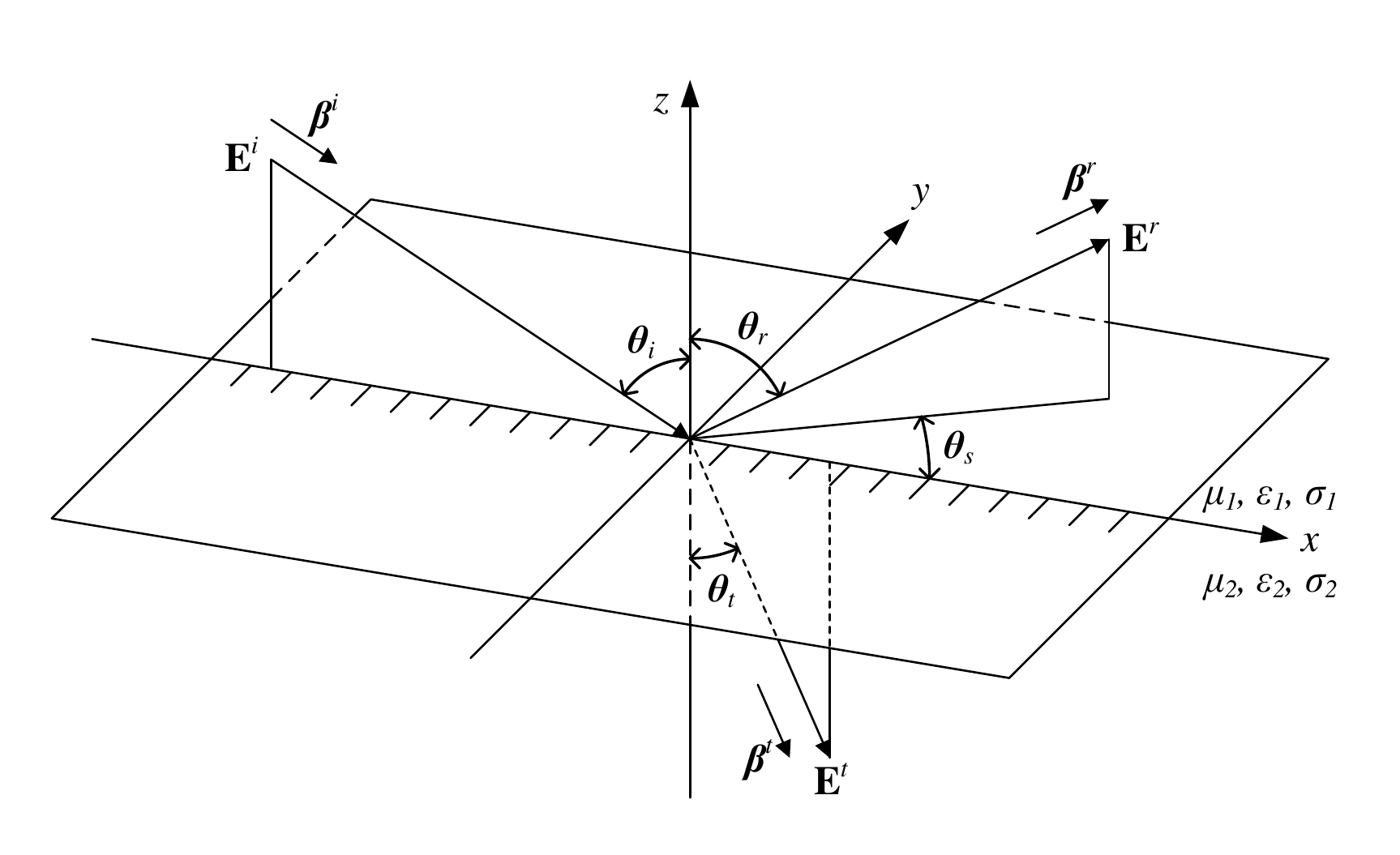}
\caption{Oblique uniform plane electromagnetic wave incidence on an interface.}
\label{fig:model:geometry}
\end{figure}

The non-line-of-sight (NLoS) EM wave propagation mechanisms are transmission, reflection, scattering and diffraction. When a uniform plane EM wave propagating in a near-perfect dielectric medium, such as air whose relative permittivity, $\varepsilon_r$, which is also termed as the dielectric constant in the literature, equals 1.0006, is incident upon a lossy medium, as shown in \figurename \ref{fig:model:geometry}, part of its wave intensity is transmitted into this medium, and the rest is reflected back. The ratios of the transmitted and reflected electric field components, $\mathbf{E}^t$ and $\mathbf{E}^r$, to the incident electric field component, $\mathbf{E}^i$, are termed as the transmission and reflection coefficients at the interface, $T^b$ and $\Gamma^b$, respectively. The equations for $T^b$ and $\Gamma^b$ depend on the polarization of the incident wave and expressed as
\begin{IEEEeqnarray}{rCl}
T^b_{\perp}&=&\frac{2\eta_2\cos\theta_i}{\eta_2\cos\theta_i+\eta_1\cos\theta_t}\IEEEyesnumber\IEEEyessubnumber\label{eq:tper}\\
\Gamma^b_{\perp}&=&\frac{\eta_2\cos\theta_i-\eta_1\cos\theta_t}{\eta_2\cos\theta_i+\eta_1\cos\theta_t}\IEEEyessubnumber\label{eq:rper}\\
T^b_{\parallel}&=&\frac{2\eta_2\cos\theta_i}{\eta_2\cos\theta_t+\eta_1\cos\theta_i}\IEEEyessubnumber\label{eq:tpar}\\
\Gamma^b_{\parallel}&=&\frac{\eta_2\cos\theta_t-\eta_1\cos\theta_i}{\eta_2\cos\theta_t+\eta_1\cos\theta_i}\IEEEyessubnumber\label{eq:rpar}%
\end{IEEEeqnarray}
for perpendicular $(\perp)$, or horizontal, and parallel $(\parallel)$, or vertical, polarizations, respectively, where $\theta_i$ and $\theta_t$ are the incident and transmission, or refracted, angles, and $\eta_1$ and $\eta_2$ are the intrinsic impedances of the dielectric and conductor media in ohms, respectively.

EM waves can have any general polarization and one method to attain the resulting $\mathbf{E}^t$ and $\mathbf{E}^r$ is separating the $\mathbf{E}^i$ into its $\perp$ and $\parallel$ components, calculating the $\mathbf{E}^t_{\perp}$, $\mathbf{E}^r_{\perp}$ and $\mathbf{E}^t_{\parallel}$, $\mathbf{E}^r_{\parallel}$ independently and vector summing the parted components.

For any polarization category, $\mathbf{E}^t$ in V/m can be written as
\begin{IEEEeqnarray}{rCl}
\mathbf{E}^t&=&\mathbf{E}_2e^{-\boldsymbol{\gamma}^t\boldsymbol{\cdot}\mathbf{r}}\IEEEnonumber\\
&=&\mathbf{E}_2\exp[-\gamma_2(x\sin\theta_t+z\cos\theta_t)]\IEEEnonumber\\
&=&\mathbf{E}_2\exp[-(\alpha_2+j\beta_2)(x\sin\theta_t+z\cos\theta_t)]\label{eq:etran}%
\end{IEEEeqnarray}
where
\begin{IEEEeqnarray}{rCl}
E_2&=&T^bE^i\label{eq:e2}\\
E^i&=&\mathbf{\hat{e}}_i\boldsymbol{\cdot}\mathbf{E}^i\label{eq:ei}\\
\mathbf{r}&=&\mathbf{\hat{a}}_xx+\mathbf{\hat{a}}_yy+\mathbf{\hat{a}}_zz\label{eq:posvec}%
\end{IEEEeqnarray}
$\boldsymbol{\gamma}^t$ is the propagation constant of the wave, $\mathbf{\hat{e}}_i$ is the unit vector in the direction of $\mathbf{E}^i$ and $\mathbf{r}$ is the position vector in rectangular coordinates. EM waves attenuate in all media except for the perfect dielectrics, and this effect is represented by the only real exponential in (\ref{eq:etran}), which includes the attenuation constant $\alpha$ in Np/m that is defined as
\begin{IEEEeqnarray}{rCl}
\alpha&=&w\sqrt{\mu\varepsilon}\left\{\frac{1}{2}\left[\sqrt{1+\left(\frac{\sigma}{w\varepsilon}\right)^2}-1\right]\right\}^{1/2}\label{eq:attcnst}%
\end{IEEEeqnarray}
where $w$ is the angular frequency in rad/s, $\mu$ is the permeability in H/m, $\varepsilon$ is the permittivity in F/m and $\sigma$ is the conductivity in S/m. A medium whose constitutive parameters, which are $\mu$, $\varepsilon$ and $\sigma$, depend on the frequency of the applied field is labelled as dispersive. Whereas all materials possess different levels of dispersion, the variation is typically insignificant, except for the permeabilities of ferromagnetic and ferrimagnetic, and permittivity and conductivities of dielectric materials. Therefore, for the ordinary lossy media, it can be assumed that only $w$ in (\ref{eq:attcnst}) increases with rising operation frequency, and so the $\alpha$. Consequently, transmission losses are greater for the low-THz band compared to the conventional spectra, which is also evidenced by respective measurements \cite{Piesiewicz-2007b}.

Similarly, $\mathbf{E}^r$ can be expressed as
\begin{IEEEeqnarray}{rCl}
\mathbf{E}^r&=&\mathbf{E}_1^re^{-j\boldsymbol{\beta}^r\boldsymbol{\cdot}\mathbf{r}}\IEEEnonumber\\
&=&\mathbf{E}_1^r\exp[-j\beta_1(x\sin\theta_r-z\cos\theta_r)]\label{eq:erefl}%
\end{IEEEeqnarray}
where
\begin{IEEEeqnarray}{rCl}
E_1^r&=&\Gamma^bE^i\label{eq:e1}\\
\beta&=&w\sqrt{\mu\varepsilon}\left\{\frac{1}{2}\left[\sqrt{1+\left(\frac{\sigma}{w\varepsilon}\right)^2}+1\right]\right\}^{1/2}\label{eq:phscnst}%
\end{IEEEeqnarray}
$\beta$ is the phase constant in rad/m, which is also known as the wave number and represented with $k$ in the literature, and $\theta_r$ is the reflected angle. Hence, the amplitude of $\mathbf{E}^r$ varies with $\Gamma^b$, which depends on $\eta$ and $\theta_t$, that are calculated using
\begin{IEEEeqnarray}{rCl}
\eta&=&\sqrt{\frac{jw\mu}{\sigma+jw\varepsilon}}\label{eq:intimp}\\
\gamma_1\sin\theta_i&=&\gamma_2\sin\theta_t\label{eq:snellRefr}%
\end{IEEEeqnarray}

(\ref{eq:snellRefr}) is Snell's law of refraction and it generates a complex $\theta_t$ for incidences comprising a lossy medium. Therefore, the true refracted angle, $\psi_t$, and the direction of wave travel, $\mathbf{\hat{n}}_t$, are written as \cite{Balanis-1989}
\begin{IEEEeqnarray}{rCl}
\psi_t&=&\tan^{-1}\left(\frac{\beta_1\sin\theta_i}{\alpha_2\,\mathrm{Im}(\cos\theta_t)+\beta_2\,\mathrm{Re}(\cos\theta_t)}\right)\label{eq:angTra}\\
\mathbf{\hat{n}}_t&=&\mathbf{\hat{a}}_x\sin\psi_t+\mathbf{\hat{a}}_z\cos\psi_t\label{eq:waveDir}%
\end{IEEEeqnarray}

Unlike the $\alpha$, there are many frequency dependent parameters contained within the non-linear equation of $\Gamma^b$. Thus, determining the precise effects of ascending carrier frequency on $\Gamma^b$, and so $\left\lvert\mathbf{E}^r\right\rvert$, via broad theoretic observations for different material classes is unachievable. Although very limited in the number of materials covered, there are existing studies which are based on actual EM wave measurements that report the $\Gamma^b$ of the evaluated specimens. One of these illustrate the $\Gamma^b_{\perp}$ and $\Gamma^b_{\parallel}$ of one sample of paperboard with ingrain wallpaper pasted on front and two concrete plaster examples, up to 1 THz and for two different $\theta_i$. Whereas the exact values are not released, the available graphs of $\Gamma^b_{\perp}$ and $\Gamma^b_{\parallel}$ show that even though there is fluctuation present in the curves, it is negligible for the computed frequency range and both $\theta_i$\cite{Piesiewicz-2007c}.

This result is actually expected for customary materials which are poorly dispersive. However, it cannot be concluded that the increasing operation frequency does not affect $\left\lvert\mathbf{E}^r\right\rvert$, since this also invalidates the smoothness assumption for the planar interface between the media under which (\ref{eq:erefl}) is derived. Therefore, a more comprehensive analysis incorporating rough surfaces into reflection and scattering formulations is needed.

An irregular, or rough, boundary, or interface, is described to have periodic or random variations of height from a pre-set mean plane, or smooth, boundary. The scattering coefficient, $\rho$, is defined as
\begin{IEEEeqnarray}{rCl}
\rho&=&\frac{E^s}{E^r}\label{eq:rho}%
\end{IEEEeqnarray}
where $\mathbf{E}^s$ is the scattered electric field component, and $E^r$ is assumed to be the reflection created by an $\mathbf{E}^i_{\perp}$ upon a smooth and perfectly conducting interface in the specular direction, whose conditions are
\begin{IEEEeqnarray}{rCl}
\theta_r&=&\theta_i\IEEEyesnumber\IEEEyessubnumber\label{eq:snellRefl}\\
\theta_s&=&0\IEEEyessubnumber\label{eq:longitudinalSct}%
\end{IEEEeqnarray}

(\ref{eq:snellRefl}) is Snell's law of reflection. The mean scattered power density, $\left\langle S^s\right\rangle$, in W/m\textsuperscript{2} is then expressed as
\begin{IEEEeqnarray}{rCl}
\left\langle S^s\right\rangle&=&\frac{1}{2}Y_1\left\langle E^s\overline{E^s}\right\rangle\IEEEnonumber\\
&=&\frac{1}{2}Y_1{\left\lvert E^r\right\rvert}^2\left\langle\rho\overline{\rho}\right\rangle\label{eq:powerS}%
\end{IEEEeqnarray}
where
\begin{IEEEeqnarray}{rCl}
E^r&=&\Gamma^b_{\perp}E^i\IEEEyesnumber\IEEEyessubnumber\label{eq:ereflSct}\\
&=&j\frac{2\beta_1E^ie^{j\beta_1R_0}XY\cos\theta_i}{\pi R_0}\IEEEyessubnumber\label{eq:ereflFull}%
\end{IEEEeqnarray}
$Y_1$ is the admittance of the dielectric medium in S, $R_0$ is the distance of the point of observation from the origin, which lies on the mean plane interface but not necessarily on the interface, $X$ and $Y$ are the dimensions of the rough boundary, the angular brackets $\left(\left\langle\right\rangle\right)$ indicate mean value, and the bar $\left(\,\bar{}\,\right)$ denotes complex conjugate. $E^r$ is of $\perp$ polarization because when the reflection is initiated by linearly polarized $\mathbf{E}^i$ and upon smooth interfaces, $\mathbf{E}^r$ preserve the linear polarization properties. (\ref{eq:ereflFull}) is also calculated exercising the Helmholtz integral for the assumptions given in the definition of $\rho$.

Let the interface between the media be rough in two dimensions and the surface is given by a random stationary process $\zeta(x,y)$, whose values designate the level $z$ of the surface at points $(x,y)$ on the $xy$ plane boundary. If $\zeta$ is normally distributed with zero mean value, standard deviation $\varsigma$, which represents the roughness of the surface, and correlation distance $D$, signifying the separation between two variable points for which the autocorrelation coefficient reduces to $e^{-1}$, and so the density of the irregularities, $\left\langle\rho\overline{\rho}\right\rangle$ is computed as
\begin{IEEEeqnarray}{rCl}
\left\langle\rho\overline{\rho}\right\rangle&=&e^{-g}\left(\rho_0^2+\frac{\pi D^2F^2}{A}\sum_{m=1}^\infty\frac{g^m}{m!m}e^{\frac{-(v_x^2+v_y^2)D^2}{4m}}\right)\label{eq:rhorho}%
\end{IEEEeqnarray}
where
\begin{IEEEeqnarray}{rCl}
g&=&\left[\frac{2\pi\varsigma}{\lambda}(\cos\theta_i+\cos\theta_r)\right]^2\label{eq:g}\\
\rho_0&=&\frac{\sin(v_xX)}{v_xX}\frac{\sin(v_yY)}{v_yY}\label{eq:rho0}\\
F&=&\frac{1+\cos\theta_i\cos\theta_r-\sin\theta_i\sin\theta_r\cos\theta_s}{\cos\theta_i(\cos\theta_i+\cos\theta_r)}\label{eq:factor}\\
v_x&=&\frac{2\pi}{\lambda}(\sin\theta_i-\sin\theta_r\cos\theta_s)\label{eq:vx}\\
v_y&=&\frac{-2\pi}{\lambda}(\sin\theta_r\sin\theta_s)\label{eq:vy}\\
\lambda&=&\frac{2\pi}{\beta_1}\label{eq:wavelength}%
\end{IEEEeqnarray}
$g$ is the Rayleigh roughness parameter, $\rho_0$ is the scattering coefficient of a smooth interface of area $A=XY$, $F$ is a factor from the general Kirchhoff solution for $\rho$, $v_x$ and $v_y$ are the $x$ and $y$ components of the vector $\mathbf{v}=\boldsymbol{\beta}^i-\boldsymbol{\beta}^r$, respectively, $m$ is the index introduced by an exponential series expansion step during the derivation of (\ref{eq:rhorho}), and $\lambda$ is the wavelength of the EM wave in the dielectric medium in m \cite{Beckmann-1986}.

\begin{figure}[!t]
\centering
\includegraphics[width=3.5in]{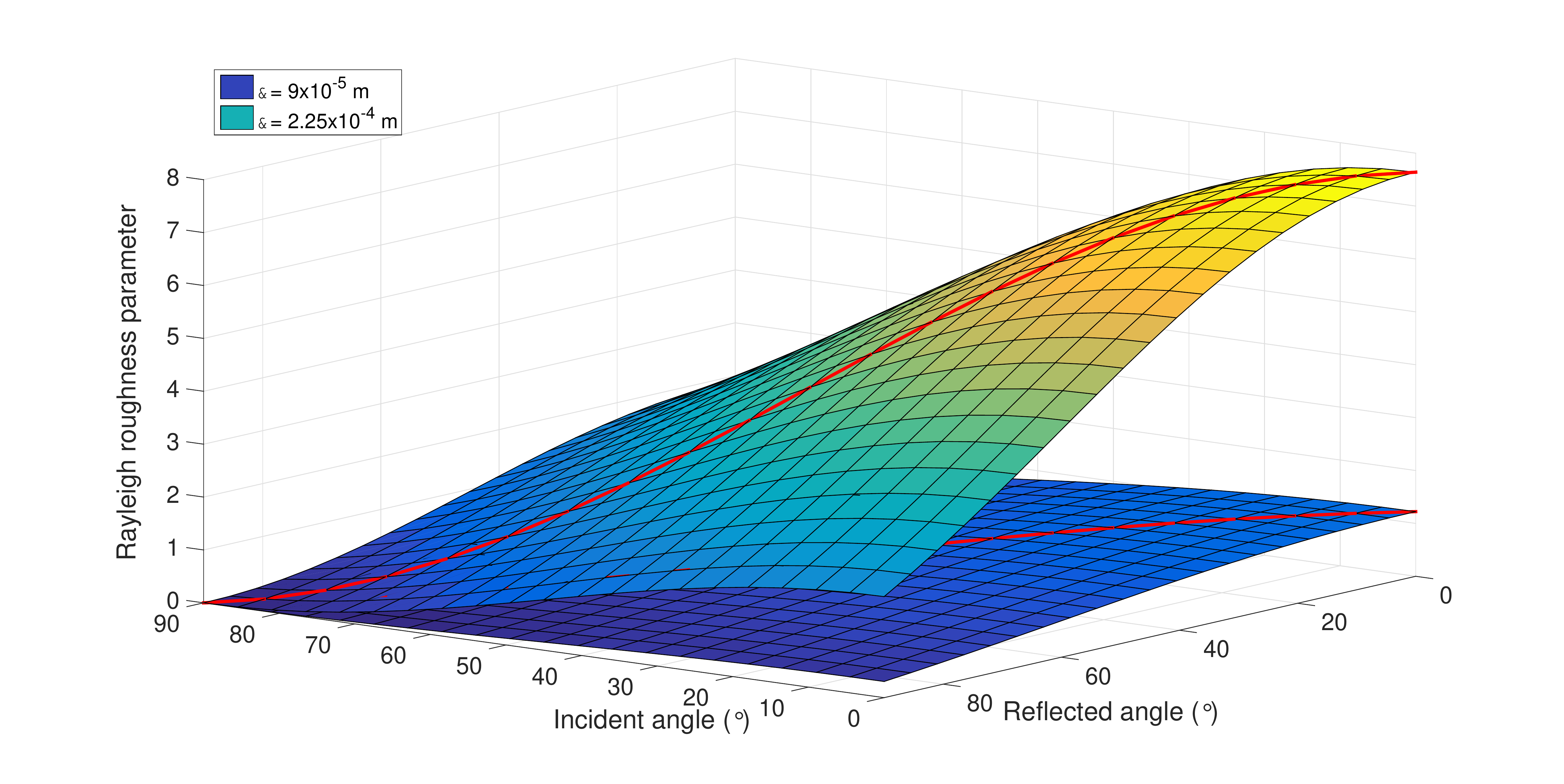}
\caption{Rayleigh roughness parameter, calculated for a uniform plane EM wave propagating in air at 293.089 GHz and for incident rough surfaces with standard deviations of 0.09 and 0.225 mm, where red lines demonstrate the cases of specular reflection.}
\label{fig:model:rrp}
\end{figure}

%

To the best of authors' knowledge, there are three publications in the literature which propose or include a channel model for the THz band. While the emphasis is on ergodic capacity calculation of a THz band communication system that utilizes hybrid beamforming on its antenna subarrays, a channel model nevertheless is available in \cite{Cen-2014}. It builds on the famous work of Saleh and Valenzuela, which primarily proposed that rays arrive in clusters in indoor transmissions. In \cite{Cen-2014}, arrival time mechanisms are kept the same as Poisson processes, whereas gain of the rays are detailed to contain gaseous attenuation. Angular characteristics and beamforming related antenna components are also included from related works. However, the model is of unknown use since its accuracy is not validated against correct channel measurements.

Of the two papers which put forward a THz band channel model, the more recent one is strictly analytical \cite{Han-2015}. Formed on the proposal to divide a wideband THz channel into narrow enough subbands that do not exhibit frequency selectivity, the channel response of each such subband is claimed to be the addition of LoS, reflection, diffraction and scattering components. Since the transfer functions and test data sets of all these propagation mechanisms are mostly based on existing work, the channel is yet to be independently analytically modelled and verified.

The only low-THz band channel model that is constructed on top of both the theoretical and experimental work of a single research group is \cite{Priebe-2013}. Outputs of an internally developed ray tracing software are used for modelling instead of real channel measurements, while omitting diffraction and scattering components. Analytical probability density functions approximating the amplitude and polarization parts of the channel transfer function are parametrized in full. However, the formulated model is scenario-specific, hence the provided time and angle of arrival and angle of departure information are of no general use and can be deterministically calculated. Altogether, while it is far from final with many issues that need to be resolved, \cite{Priebe-2013} still presents the first true low-THz band channel model in the literature.

\section{THz Band Device Technologies}
\label{sec:device}

While the THz band seems to offer an abundant spectrum for every radio service conceivable, it actually presents a very harsh environment for EM wave propagation. Although prohibitively high attenuation by atmospheric gases is advantageous for some small number of specific applications, like intersatellite communications links because it assists those to be isolated from any possible interference from the Earth, THz band is yet to be utilized for communication purposes. However, this has not been the case for the whole scientific field as the THz band contains some unique and valuable information that are within the research interests of different areas. For example, temperatures of the interstellar dust clouds range between 10 and 200$^{\circ}$K, which corresponds to about 210 GHz and 4.3 THz, respectively. Therefore, energy radiated from interstellar gas, which is used for star formation research, lies entirely within the THz band. Gases that make up the Earth's atmosphere also have thermal emission lines in the THz band, creating Earth science's need for measurement instruments working at THz frequencies.

Several device technologies are available today which jointly cover the transmitter (TX) and receiver (RX) needs of the entire THz spectrum, ranging from metamorphic high electron mobility transistors to quantum cascade lasers; however, only a very small percentage of these possess the potential to be used for 5G communication systems. By 2020, 11.6 billion mobile-connected devices are expected to be in use \cite{ciscoVNI-2015}. If 5G systems are to acquire high market penetration rates, respective UE and network devices must be robust, lightweight, highly integrated and most importantly, low-cost. Taking into account the technologies which are currently used to manufacture the hardware of virtually all mainstream communication devices and after a review of the currently available THz device technologies, silicon (Si) and complementary metal-oxide-semiconductor (CMOS) appear as the only viable candidates for 5G despite their shortcomings in practically all electronic performance criteria.

When compared to the III-V semiconductor compounds, Si has worse material properties than many. Lower electron mobility, smaller energy band gap and higher resistivity of Si results in devices with inferior figures of merit. CMOS, likewise, has poorer transistor and passive performances than corresponding components produced using III-V compound processes. However, there are plenty of solid reasons that have caused Si and CMOS dominate the global semiconductor market, and with the current development rate in corresponding areas, their places look secure. Si, first of all, is vastly available all around the world and its purification is simple. Mechanical characteristics of Si make it a sturdy material, thus easy to manufacture and handle. Si also has high thermal conductivity enabling efficient thermal management of devices. It is simple to form insulators with exceptional dielectric properties like silicon dioxide that are used as CMOS transistor gates among many other functions. The doping concentration of Si has a very high range and with the already established manufacturing capacity and continued demand, low-cost production is ensured. On the other hand, since CMOS technology is essentially the same for all devices regardless of the frequency of operation, CMOS's intrinsic advantages like integration of higher frequency circuits with baseband circuitry, digital calibration for better performance, high yield and built-in self test also holds true for the THz range devices.

Considering the May 2014 start of TG3d for 300 GHz standardization, it will be safe to say we are at least a decade away from commercial low-THz band products. Nevertheless, research activities on both circuitry and communication domains are starting to pick up speed. Emerging THz band applications which also develop Si CMOS technology include imaging \cite{Uzunkol-2013}, sensors \cite{Karolyi-2014} and chip interconnection \cite{Gu-2015}. Artificial dielectric layers are proposed to improve performance of on-chip antennas radiating at low-THz band \cite{Syed-2015}. However, in line with the subject of the paper, in the following subsections a complete survey is presented on the state-of-the-art Si CMOS THz circuit blocks and modules designed for communication purposes. Moreover, devices whose operation frequencies are contained by the first three transmission windows within the low-THz band, which, approximately, range from 275 to 420 GHz, are selected in order to demonstrate the potential for 5G in the THz region. This limitation imposed the exclusion of a number of notable Si CMOS studies that are just outside the frequency range \cite{Ruonan-2013, Yen-Ju-2013}.

\subsection{Signal Sources}
\label{sec:device:source}

THz signal source fabrication using CMOS is probably the most difficult field of the THz electronics research and only recently have implementations with acceptable power, high stability and frequency tuning been published. Depending on the power gain cutoff frequency ($f_{max}$) of the transistors that are used, low-THz band can be reached through either frequency multiplication or direct generation. If the $f_{max}$ of a device is large enough for the intended THz application, direct generation is commonly preferred since power efficiency is better and smaller chip area is needed compared to frequency multiplication. However, especially for Si CMOS devices, this is mostly not the case. Therefore, signal sources are specifically designed to efficiently generate power at the harmonic frequencies of built-in non-linear diodes, so that appropriate harmonics of the fundamental frequency ($f_0$) can be output.

One Si CMOS source employing triple-push architecture is reported in \cite{Grzyb-2013}. An N-push oscillator consists of N coupled oscillators which use a shared resonator and output $2\pi/N$ phase-shifted signals. When these signals are combined, N\textsuperscript{th} harmonic components are constructively added, whereas the rest, in theory, are negated. While this method is useful for higher frequency generation, discontinuous tuning is observed in the event of uneven phase-shift \cite{Rohde-2005}.

In \cite{Grzyb-2013}, same two triple-push oscillator cores are locked by magnetic coupling, and the power is conveyed to the differential ring antenna through a matching stage. The device is realized in a 65-nm CMOS process over a 500 $\times$ 570 $\mu$m\textsuperscript{2} die area, oscillators occupying 120 $\times$ 150 $\mu$m\textsuperscript{2} of the total. Output frequency is measured to be tunable from 284 to 288 GHz by reducing the supply voltage from 1.4 to 0.7 Volts (V), and the source can generate a peak output power of -1.5 dBm at the upper limit of the tuning range by consuming 275 milliwatts (mW) of DC power. The circuitry was also packaged with a Si lens on an FR-4 board, but since the aim was demonstration and did not involve original design, that part is omitted.

Another novel CMOS source that is implemented in a 65-nm low-power bulk CMOS process is presented in \cite{Tousi-2012}. For frequency tuning, placing varactors inside the LC resonator is a common practice that is shown to work satisfactorily up to 0.1 THz \cite{Shinwon-2014}. However, at higher frequencies varactor performance degrades. The significance of the design of \cite{Tousi-2012} originates from eliminating varactors from the voltage-controlled oscillator (VCO), but still delivering a frequency tunable source with high output power which functions at the beginning of the submm-wave band. By adding phase shifters to the proposed four core coupled oscillator system, locking frequency of the VCO is made adjustable in accordance with the phase shifts between each core and the respective injected signal. Even though the provided simulation result illustrated that the third harmonic generates higher current around 300 GHz, fourth harmonic is chosen also for the symmetry it brings. One of the two VCOs that are fabricated for the study radiates peak output power of -1.19 dBm around the 13 GHz of tuning range centred at 290 GHz, therefore achieving the highest output power and tunability for all the oscillators available in the literature which operate in and beyond the low-THz band, even including ones using compound semiconductor technologies. DC-to-RF conversion efficiency stands at 0.23\% due to the 325 mW DC power input, and the chip is printed on an area of 600 $\times$ 600 $\mu$m\textsuperscript{2}.

\subsection{Transmitters}
\label{sec:device:tx}

Not just sources but also complete TXs are being developed for THz frequencies. The latest such device is a phased array \cite{Tousi-2015} that is expanded over the delay-coupled oscillator method by the authors of \cite{Tousi-2012}. The idea of controlling the oscillator frequency through the phase shift between adjacent cores works on a one-dimensional ring. To extend this effect over two dimensions, in \cite{Tousi-2015} a 2 $\times$ 2 central loop is connected to 4 other similar loops only through one of its vertices, creating a 4 $\times$ 4 coupled array. Adjacent nodes are situated at a fixed distance that equals half of the radiation wavelength, and the oscillators are linked with phase shifters. Patch antennas are used for radiation to prevent substrate coupling. A sample, likewise \cite{Tousi-2012}, is manufactured in a 65-nm bulk CMOS process over an area of 1.95 $\times$ 2 mm\textsuperscript{2}. Peak total radiated power is measured at 338 GHz as 0.8 mW, or -0.97 dBm, and equivalent isotropically radiated power (EIRP) as 51 mW, or 17.1 dBm, using 1.54 W. 12 dB of the 18 dB antenna directivity is due to array gain, and the rest to patch antenna directivity. Beam steering is feasible across 45$^{\circ}$ in azimuth ($\Phi$), and 50$^{\circ}$ in elevation ($\Theta$) angles. Centre frequency tuning measurements are performed between 337 and 339 GHz. However, in the paper 2.1\% tuning is claimed to be possible via altering the coupler supply voltage, which results in a 7.1 GHz spectrum around 338 GHz.

One other architecture is tunable at second harmonic frequencies that are between 276 and 285 GHz \cite{Sengupta-2012}. Distributed active radiator (DAR) \cite{Sengupta-2011} and inverse design approach lies at the core of the design. Typically, power generation and radiation are implemented by different circuit blocks. However, in \cite{Sengupta-2012}, surface currents on Si chip metal layers are first calculated for a specific EM field, and then synthesized using a DAR, which is made of four cross-coupled transistor pairs located symmetrically along two loops that are shaped into a M\"obius strip. This way, second harmonic signal is radiated, while fundamental and other harmonic signals are filtered. The implementation consists of 16 DAR cores in a 4 $\times$ 4 array, and it is realized in a 45-nm CMOS silicon on insulator process. The output of the center VCO, which is tunable from 91.8 to 96.5 GHz using a 1.1 V supply voltage, is distributed to four separate divide-by-two frequency dividers that generate quadrature in-phase (I) and quadrature (Q) signals. Signal is then transferred through phase rotator and frequency triplers to drive the DARs. The resulting circuit, which has a chip area of 2.7 $\times$ 2.7 mm\textsuperscript{2}, is capable of beam steering nearly 80$^{\circ}$ in both $\Theta$ and $\Phi$ planes and provides an EIRP of 9.4 dBm at 280.8 GHz with the help of 16 dBi maximum directivity. 

\subsection{Transceivers}
\label{sec:device:trx}

The final integrated transceiver (TRX) model \cite{Jung-Dong-2012} is included to provide a TRX example, even though the device is implemented in 130-nm silicon-germanium bipolar CMOS process. Moreover, its 367 to 382 GHz working range is around an atmospheric attenuation local maximum, thus making the device unsuitable for communication purposes. The TRX design is a homodyne frequency-modulated continuous-wave radar using triangular modulation signal. Differential Colpitts VCO generates the fundamental signal at 92.7 GHz with 8.3\% tuning radius, which is followed by drive amplifiers. Inside the TX, initially, balanced quadrature I and Q signals are coupled through transformers to the frequency quadrupler. Two push-push pairs compose the quadrupler, which outputs the fourth harmonic frequency and separate patch antennas, each containing two patches, radiates and receives the signal. On the RX side a subharmonic mixer, driven by second harmonic quadrature I and Q signals, down-converts to intermediate frequency (IF), before the concluding IF amplifier stage. The TX translates 3 dBm VCO output power into -14 to -11 dBm EIRP, the RX noise figure is assessed to be between 35 and 38 dB, and the entire TRX consumes 380 mW power above a total space of 2.2 $\times$ 1.9 mm\textsuperscript{2}.

\section{Open Research Issues}
\label{sec:iss}

Investigations on low-THz band for wireless communication purposes began nearly a decade ago. Consequently, the research area is only emergent with several problems which need to be solved before real world deployments. The foremost of these can be described as follows:
\begin{itemize}
\item \textbf{Realistic channel models:} As previously explained, for low-THz band there is only one channel model candidate that incorporates serious inadequacies. Multiple new models which target potential WPAN and WLAN use cases should be devised based on actual channel measurements performed in various environments including outdoors, where the effects of meteorological phenomena also ought to be accounted for. In addition to providing a complete and accurate elucidation of the small- and large-scale space-time-frequency characteristics of the channel, the models need to be adaptable to analytical and empirical analyses for higher communication layer research subjects.
\item \textbf{Multiple channel access:} The very high data rate and zero-latency requirements from the low-THz band for IoT applications are achievable through multihop communication over dense ad hoc networks, and the multiple access scheme should be compliant. Multiple, hybrid and random access methods need to be investigated for varying channel bandwidth, equipment density and usage models. While intersymbol interference is probably insignificant since multipath is confined in the THz band, results of multiple access interference ought to be examined for different access techniques.
\item \textbf{Baseband signal processing:} For successful low-THz band system realizations, TRX design and communication methods are equally important. However, baseband processing of a 40/100 Gb/s wireless link solely in the digital domain is a major challenge in its own right. Analogue implementations of various TRX components should be investigated as mixed-signal baseband operation can theoretically result in simpler, faster and more energy efficient circuitries. Another research area is parallel data transmission using OFDM or parallel sequence spread spectrum, because utilizing multiple streams would lower the performance requirements on the baseband processor in return for increased number of circuit elements.
\item \textbf{Multiple antenna systems:} Since the effective area of an antenna is inversely proportional to the square of signal frequency, advanced multiple antenna methods are both critical in augmenting the link budget of and practicable for low-THz band IoT services. Beam forming and steering algorithms for phased array antennas are among the prioritized investigation directions as those will aid the problems of device discovery and propagation in susceptible THz channels. Massive MIMO also necessitates further research to resolve predicaments such as pilot contamination and reciprocity calibration prior to becoming an enabling technology.
\item \textbf{Access network architecture:} Additional losses intrinsic to the THz band limit the range of individual devices. However, network densification which accompanies IoT and expanding universal Internet access present new design opportunities for uninterrupted connectivity. Last-mile network architectures should be capable of supporting the expected escalating capacity needs of each small cell, where wavelength-division multiplexing passive optical networks can become the bottleneck. Moreover, short-range communication techniques for the low-THz band need to be researched as complementary device-to-device links are effective in avoiding brief shadowing events like human blockage.
\end{itemize}

\section{Conclusion}
\label{sec:con}
With no end of wireless communication upsurge in sight, neither throughput nor data rate requirements of 5G systems can be provided by just concentrating on spectral efficiency solutions. Due to the constantly improving low-cost device technologies, the first transmission window within the THz band is no longer out of reach of widespread communication systems. As the THz standardization activities for WPANs recently progressed to second stage, research efforts in the area are to intensify even more. While the initially arising difficulties for low-THz indoor access network architecture can be resolved using the already laid out concepts, the proposed frequency leap is not small and the work has just begun. Realistic channel models, optimum multiple access algorithms and Si CMOS devices all separately yet simultaneously need development. However, the question is not the possibility of commonplace low-THz band communications, but the timetable.

\section*{Acknowledgement}
This work was supported in part by the Scientific and Technological Research Council of Turkey (TUBITAK) under grant \#113E962.

\bibliographystyle{IEEEtran}


\begin{IEEEbiography}[{\includegraphics[width=1in,height=1.25in,clip,keepaspectratio]{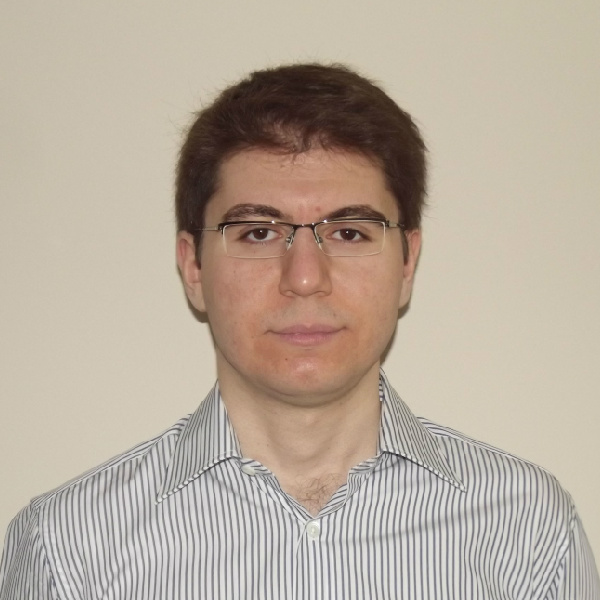}}]{Turker Yilmaz}
(S'13) received B.S. and MSc degrees in electrical and electronics engineering from the Bogazici University and University College London in 2008 and 2009, respectively. He is currently a research assistant at the Next-generation and Wireless Communications Laboratory and pursuing his Ph.D. degree within the Department of Electrical and Electronics Engineering, Koc University, Istanbul, Turkey. His current research interests include terahertz communications and Internet of Things.
\end{IEEEbiography}


\begin{IEEEbiography}[{\includegraphics[width=1in,height=1.25in,clip,keepaspectratio]{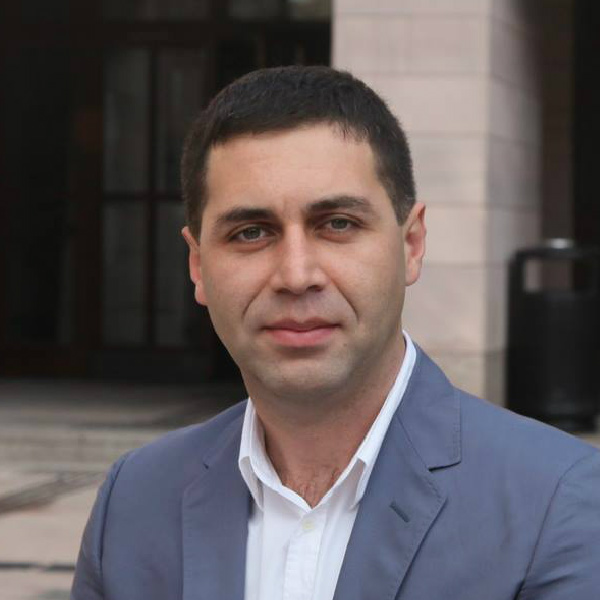}}]{Ozgur Baris Akan}
(M'00-SM'07-F'16) received Ph.D. degree in electrical and computer engineering from the Broadband and Wireless Networking Laboratory, School of Electrical and Computer Engineering, Georgia Institute of Technology, Atlanta, in 2004. He is currently a full professor with the Department of Electrical and Electronics Engineering, Koc University, the Director of the Graduate School of Sciences and Engineering, Koc University, and the Director of the Next-generation and Wireless Communications Laboratory (NWCL).

His current research interests are in nanoscale, molecular communications, next-generation wireless communications, Internet of Things, 5G mobile networks, sensor networks, distributed social sensing, satellite and space communications, signal processing, and information theory. He is an associate editor for the IEEE Transactions on Communications, IEEE Transactions on Vehicular Technology, and IET Communications, and editor for the International Journal of Communication Systems (Wiley), European Transactions on Telecommunications, and Nano Communication Networks Journal (Elsevier).
\end{IEEEbiography}

\end{document}